\documentclass[journal,screen]{acmart}
\usepackage{multirow}
\usepackage{svg}
\AtBeginDocument{%
  }

\begin{document}

%%
%% The "title" command has an optional parameter,
%% allowing the author to define a "short title" to be used in page headers.
\title{An Empirical Study of Bitwise Operators Intuitiveness through Performance Metrics}

%%
%% The "author" command and its associated commands are used to define
%% the authors and their affiliations.
%% Of note is the shared affiliation of the first two authors, and the
%% "authornote" and "authornotemark" commands
%% used to denote shared contribution to the research.
\author{Shubham Joshi}
\email{joshis4@unlv.nevada.edu}
\orcid{0000−0001−5584−3265}
\affiliation{%
  \institution{University of Nevada Las Vegas}
  \city{Las Vegas}
  \state{Nevada}
  \country{USA}
}

%%
%% By default, the full list of authors will be used in the page
%% headers. Often, this list is too long, and will overlap
%% other information printed in the page headers. This command allows
%% the author to define a more concise list
%% of authors' names for this purpose.
\renewcommand{\shortauthors}{Joshi S.}

%%
%% The abstract is a short summary of the work to be presented in the
%% article.
\begin{abstract}
    \textbf{Objectives }This study aims to investigate the readability and understandability of bitwise operators in programming, with the main hypothesis that there will be a difference in the performance metrics (response time and error rate) between participants exposed to various bitwise operators related questions and those who are not.
    
    \noindent \textbf{Participants }
    Participants in this human research study include people without programming background, novice programmers, and university students with varying programming experience (from freshmen to PhD level). There were 23 participants in this study.
    
    \noindent \textbf{Study Methods } This study uses a within-subjects experimental design to assess how people with diverse programming backgrounds understand and use bitwise operators. Participants complete tasks in a JavaScript program, and their task completion times and task accuracy are recorded for analysis.
    
    \noindent \textbf{Findings } The results indicate that operators can be one of the factors predicting response time, showing a small but significant effect (R-squared = 0.032, F(1, 494) = 16.5, p < .001). Additionally, operators such as OR, NOT, and Left Shift showed statistical significance in task completion times compared to other operators.
    
    \noindent \textbf{Conclusions } While the complexity of bitwise operators did not generally result in longer task completion times, certain operators were found to be less intuitive, suggesting the need for further investigation and potential redesign for improved understandability.
\end{abstract}

%%
%% The code below is generated by the tool at http://dl.acm.org/ccs.cfm.
%% Please copy and paste the code instead of the example below.
%%

%%
%% Keywords. The author(s) should pick words that accurately describe
%% the work being presented. Separate the keywords with commas.

% \received{20 February 2007}
% \received[revised]{12 March 2009}
% \received[accepted]{5 June 2009}

%%
%% This command processes the author and affiliation and title
%% information and builds the first part of the formatted document.
\maketitle

\section{Introduction}
Bitwise operators are an integral part of many programming languages, allowing programmers to manipulate data at the bit level. Many common operators function on either single or multiple bytes, typically containing eight bits in most systems. While not all programming languages accommodate bitwise operators, notable examples include C, Java, JavaScript, Python, and Visual Basic. By enabling finer precision and requiring fewer resources, bitwise operators can enhance code performance and efficiency. Applications of bitwise operations span various domains, including encryption, compression, graphics, communication via ports and sockets, embedded systems programming, and finite state machines.

However, the use of existing symbols for these operators can pose significant challenges for programmers, particularly those new to the field. Much of the syntax in programming languages lacks intuitiveness, as it is designed without evidence-based standards. This lack of intuitiveness often necessitates additional learning time and cognitive effort, hindering overall efficiency and comprehension of code.

Despite their widespread usage, limited research has been conducted to empirically evaluate the intuitiveness of bitwise operators from a human-computer interaction perspective. Existing studies have primarily focused on the computational aspects and performance optimization achieved through these operators, overlooking their impact on programmer experience and code readability.

This study aims to bridge this gap by employing an empirical approach to assess the intuitiveness of bitwise operators through key performance metrics, such as task completion time and error rates. By analyzing participants’ responses to tasks involving various bitwise operations, this research seeks to identify potential areas of improvement and inform the design of more intuitive programming language constructs, ultimately enhancing developer productivity and code quality.

\subsection{Problem}
The use of bitwise operators in programming languages can be challenging for new users due to the unintuitive symbols and their functions. The lack of self-explanatory symbols often leads to confusion, requiring additional learning time and cognitive effort as users must decipher the meaning and purpose of these operators. Improving the clarity and accessibility of bitwise operators is crucial for enhancing code readability and comprehension, which can ultimately streamline the software development process.

\paragraph{RQ} How intuitive are bitwise operators in general?

\subsection{Review of Relevant Scholarship}

The purpose of reviewing relevant research is to provide a comprehensive overview of existing literature related to general programming languages and their syntax, as well as their intuitiveness. We will explore studies peripheral to the main focus of our research.

Weidmann, Theo B. (2022)~\cite{weidmann2022} introduces a new programming language based on a programming-by-demonstration paradigm. This enables users to experiment with and test programs, describe complex operations without syntax, and display an approximation of the program state while coding. Researchers in the literature aim to provide a better programming experience for developers overall. Similarly, Denny, Paul et al. (2011)~\cite{denny2011} investigate the frequency of syntax errors encountered by students during a drill and practice activity. They aim to understand the challenges novice programmers face in mastering syntax, particularly when writing short fragments of code. The study sheds light on the extent of students' struggles with syntax, revealing potential areas for improvement in programming education.

Gilsing, M. (2022)~\cite{gilsing2022} designed and implemented the Hedy programming language following a gradual approach to lower the syntax barrier for beginners. They conducted an empirical user study involving 39 children aged 11 to 14, assessing the impact of Hedy’s gradual nature on learning and highlighting both its advantages and challenges in terms of syntax usability.

Stefik and Gellenbeck (2011)~\cite{stefik2011} conducted two empirical studies investigating how programmers interact with their development environment. Their research indicates that while sound cues might improve debugging efficiency with practice, novice programmers find the terminology of common programming languages unintuitive. Stefik and Siebert (2013)~\cite{stefik2013} expanded upon their earlier work on the intuitiveness of programming languages. Their four new studies continued to explore how novices understood both individual constructs and larger structural elements such as loop structures. The findings offer insights for instructors and language designers, emphasizing the importance of syntax in novice programming education and suggesting strategies for improvement.

Additionally, there are studies discussing intuitiveness in programming languages. Stefik et al. (2011)~\cite{stefik2011ieee} explored the possibility of creating a programming language named Hop, where syntax, semantics, and API design are based on rigorous empirical data. They addressed the historical reliance on expert opinions in programming language design, advocating a more scientific approach. Similarly to what our paper proposes, an experimental study needs to be conducted to implement elements in a programming language intuitively. In one study, Diep, M. and Cheimonettos, A.~\cite{diep2023} developed a set of guidelines for designing intuitive query languages that minimize cognitive load for users.

In addition, researchers have focused on improving program readability and developer productivity by addressing issues in programming languages. Deißenbock (2005)~\cite{deißenbock2005} proposed a solution to the challenge of identifier naming in programming languages. Their approach improves program readability, promotes consistent naming practices, and enhances programmer productivity. Similarly, other researchers have worked on different topics in the realm of programming languages, such as Høst (2007)~\cite{høst2007}, who focused on naming methods, and Binkley (2009)~\cite{binkley2009}, who investigated the impact of identifier style (specifically camel casing versus underscores) on code readability.

Existing literature shows the efficiency and speed of operations leveraging bitwise operators (Manohar, 2015)~\cite{manohar2015}. Bitwise operations can improve the efficiency and speed of an algorithm by directly manipulating data at the bit level (Yordzhev, 2012)~\cite{yordzhev2012}. However, there remains a gap in research regarding the impact of bitwise operators on readability and understandability for developers or anyone using bitwise operators. There is literature where computer scientists have contributed to improving programming languages. Yet, our goal is to identify whether currently used bitwise operators are intuitive and to suggest alternatives based on empirical methods.
\subsection{Hypothesis, Aims, and Objectives}
Following are the hypothesis for this study:

\paragraph {Hypothesis 1}
There is no linear association between operators and task completion times.

\paragraph {Hypothesis 2}
The complexity of bitwise operations, including AND, OR, XOR, Left Shift, Right Shift, does not result in longer task completion times.

\section{Method}

\subsection{Inclusion and Exclusion}
Inclusion criteria for this study encompass individuals with no programming language experience, as well as those with limited programming experience, referred to as novices. In addition, participants from various academic levels in university settings were considered. There were no limitations based on demographic characteristics such as gender or ethnicity. Exclusion criteria for this study include individuals who are legally barred from participating, such as children and prisoners.

\subsection{Participant Characteristics}
Participants with a computer science background were considered for the study. The age of the participants ranged from 19 to 31 years. Their programming experience ranged from 0 to more than 5 years.

\subsection{Sampling Procedures}
Convenience sampling was used for participant selection, allowing individuals to participate based on accessibility and willingness. This method, while lacking the randomness of other sampling techniques, facilitated the inclusion of participants who were readily available and interested in engaging with the study. Self-selection may have occurred because participants voluntarily accessed the study website and followed the instructions provided; however, no specific data on self-selection rates were collected. All individuals who agreed to participate were included in the study, with no systematic sampling plan implemented.

Data were collected asynchronously through a dedicated website, accessible from any location with internet connectivity. Participants were required to visit the website, read instructions, and proceed with the study activities. The specific locations of the participants were not tracked or recorded. Participants engaged in the study voluntarily, providing consent through verbal and written agreements. No monetary payments or incentives were offered for participation. Although no approval was obtained from an institutional review board prior to the study, ethical standards were maintained throughout the research process.

\subsection{Sample Size, Power and Precision}
In the literature, there are not many studies specifically focused on bitwise operators to compare effect sizes. For this reason, it is difficult to determine an appropriate sample size for this study.

\subsection{Measures and Covariates}
The primary measures in this study focused on participant performance in tasks involving bitwise operators. These measures included the accuracy of completed tasks, the time taken to complete each individual task, and the speed and precision of responses in experiments with multiple-choice questions. Collected demographic variables included participant age, gender, academic status, and years of professional experience.

\subsection{Data Collection}
The study used a custom-built JavaScript web application to collect data. The application included both demographic surveys (age, programming experience, academic status, gender) and experimental tasks focused on bitwise operators. Data collection could take place either in a controlled onsite setting or remotely, according to participants' convenience. The experimental results were stored in a CSV file for later statistical analysis. Table 1 provides an example of the data format collected. The application was accessible through the dedicated website exp.shjoshi.com.np, and a screenshot of the website is provided in Figure 1.

Each task was associated with an operator. During data collection, the accuracy and response time for each operator were named as shown in Table 2.

\begin{table}[htbp]
\centering
\begin{tabular}{|c|c|c|}
\hline
Operator & Accuracy Variable & Response Time Variable \\
\hline
\multirow{2}{*}
\text{$\&$} & acc\_and1 & res\_and1 \\
\hline
\texttt{*} & acc\_and2 & res\_and2 \\
\hline
\texttt{|} & acc\_or1 & res\_or1 \\
\hline
\texttt{and} & acc\_and3 & res\_and3 \\
\hline
\texttt{U} & acc\_placebo1 & res\_placebo1 \\
\hline
\texttt{or} & acc\_or2 & res\_or2 \\
\hline
\texttt{xor} & acc\_xor1 & res\_xor1 \\
\hline
\texttt{$\wedge$} & acc\_xor2 & res\_xor2 \\
\hline
\texttt{||} & acc\_or3 & res\_or3 \\
\hline
\texttt{ExclusiveOR} & acc\_xor3 & res\_xor3 \\
\hline
\texttt{@} & acc\_placebo2 & res\_placebo2 \\
\hline
\texttt{-} & acc\_not1 & res\_not1 \\
\hline
\texttt{<-} & acc\_lshift1 & res\_lshift1 \\
\hline
\texttt{$<<$} & acc\_lshift2 & res\_lshift2 \\
\hline
\texttt{\textasciitilde} & acc\_not2 & res\_not2 \\
\hline
\texttt{\&\&} & acc\_and4 & res\_and4 \\
\hline
\texttt{rshift} & acc\_rshift1 & res\_rshift1 \\
\hline
\texttt{not} & acc\_not3 & res\_not3 \\
\hline
\texttt{$\wedge\wedge$} & acc\_xor4 & res\_xor4 \\
\hline
\texttt{<\$} & acc\_lshift3 & res\_lshift3 \\
\hline
\texttt{lshift} & acc\_lshift4 & res\_lshift4 \\
\hline
\texttt{O} & acc\_palcebo3 & res\_palcebo3 \\
\hline
\texttt{!} & acc\_not4 & res\_not4 \\
\hline
\texttt{$>>$} & acc\_rshift2 & res\_rshift2 \\
\hline
\texttt{>\$} & acc\_rshift3 & res\_rshift3 \\
\hline
\texttt{$\cdot$} & acc\_and5 & res\_and5 \\
\hline
\texttt{->} & acc\_rshift4 & res\_rshift4 \\
\hline
\end{tabular}
\caption{Descriptions of Symbols with Accuracy and Response Time Variables}
\label{tab:symbols_accuracy_response}
\end{table}

\begin{figure}
    \centering
    \includegraphics[width=8cm]{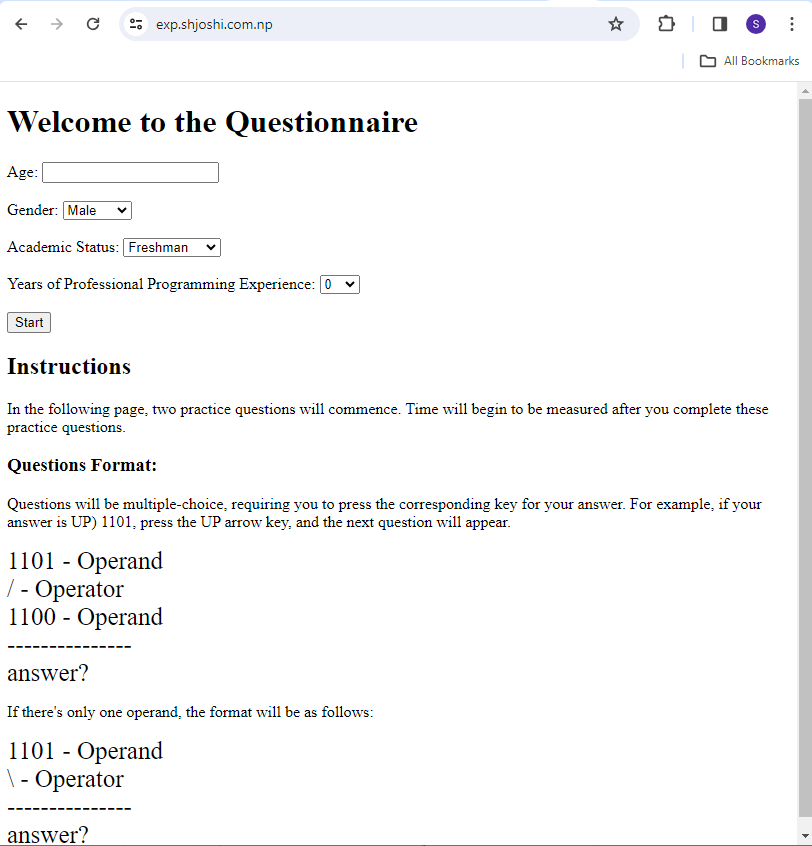}
    \caption{Webiste for the Experiment}
    \Description{Screenshot showing the experimental website interface, with navigation controls and example task interface for study participants.}
    \label{fig:website}
\end{figure}

\begin{table}[htbp]
  \centering
  \caption{Sample Data from the Dataset}
    \begin{tabular}{ccccccc}
    \toprule
    participant & age  & gender & academic\_status & prof\_exp\_years & acc\_and1 & res\_and1 \\
    \midrule
    1     & 29    & Male  & Professional & 4     & 1     & 3240 \\
    2     & 27    & Male  & Professional & 5+    & 0     & 3266 \\
    3     & 23    & Male  & Master & 1     & 0     & 1986 \\
    4     & 27    & Female & Master & 3     & 1     & 15904 \\
    5     & 27    & Male  & Master & 2     & 1     & 10667 \\
    6     & 27    & Male  & Master & 3     & 0     & 318 \\
    7     & 28    & Male  & PhD   & 0     & 1     & 8240 \\
    8     & 27    & Female & Master & 0     & 1     & 7471 \\
    \bottomrule
    \end{tabular}%
  \label{tab:addlabel}%
\end{table}%

\subsection{Quality of Measurements}
Since the experiment recorded data directly through the program, we aimed to eliminate most human factors to ensure data reliability. Participants used the 'Up Arrow Key' and 'Down Arrow Key' to select answers rather than a mouse. This approach helped improve accuracy and reduce factors that could impact response time. Additionally, two practice questions were provided before the experiment to familiarize participants with the interface and controls.

\subsection{Instrumentation}
The instruments used in this study were ad hoc, consisting of experimental tasks developed with the JavaScript programming language specifically designed for the purposes of this research. Figure 1 shows the website that was used to present the experiments to the participants.

\subsection{Masking}
In this experimental study, no masking was implemented. Therefore, no interpretation regarding masking is provided in this experiment.

\subsection{Psychometrics}
As the experiment was conducted with a group of participants who were required to complete the experiment only once, no reliability tests such as test-retest were conducted.

\subsection{Conditions and Design}
To verify the hypothesis, the experiment was designed to comprise 27 questions. It consisted of 5 Bitwise AND-related questions, 3 Bitwise OR-related questions, 4 Bitwise XOR-related questions, 3 Bitwise NOT-related questions, 4 Bitwise Left Shift-related questions, and 4 Bitwise Right Shift-related questions. The sample questions asked in the experiment are shown in Figure 1. To review the complete set of operators used in the experiment, refer to Table 1, which lists all the operators included.

\begin{figure}
    \centering
    \includegraphics[width=8cm]{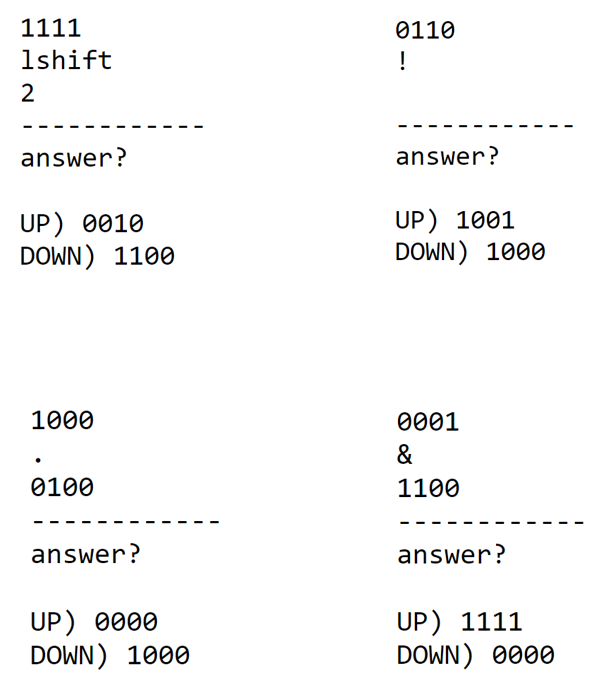}
    \caption{Sample Questions of the Experiment}
    \Description{This image displays example questions that were presented to participants during the experiment, covering various bitwise operator concepts.}

    \label{fig:questions}
\end{figure}

\subsection{Data Diagnostics}
Performance responses are recorded in CSV format in order to later analyze the data with statistical approaches and to represent them in relevant charts. The data for various operators are shown in the boxplots in Figure 3, Figure 4, Figure 5, Figure 6, Figure 7, and Figure 8.

\begin{figure}
    \centering
    \includesvg[width=8cm]{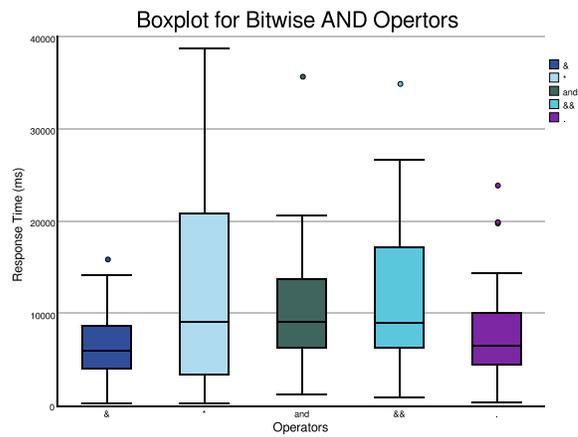}
    \caption{Boxplot Bitwise AND}
    \label{fig:andboxplot}
\end{figure}
\begin{figure}
    \centering
    \includesvg[width=8cm]{images/orboxplot.svg}
    \caption{Boxplot Bitwise OR}
    \label{fig:orboxplot}
\end{figure}
\begin{figure}
    \centering
    \includesvg[width=8cm]{images/xorboxplot.svg}
    \caption{Boxplot Bitwise XOR}
    \label{fig:xorboxplot}
\end{figure}
\begin{figure}
    \centering
    \includesvg[width=8cm]{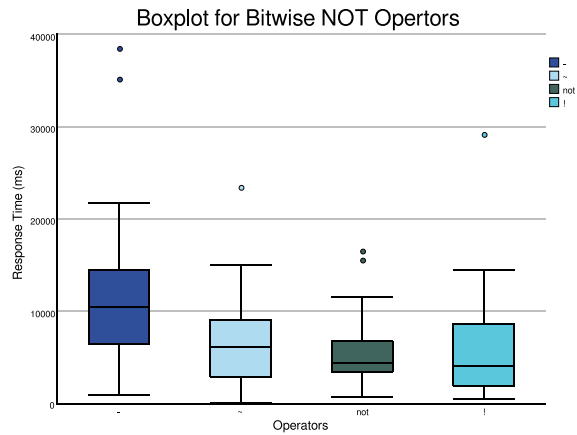}
    \caption{Boxplot Bitwise NOT}
    \label{fig:notboxplot}
\end{figure}
\begin{figure}
    \centering
    \includesvg[width=8cm]{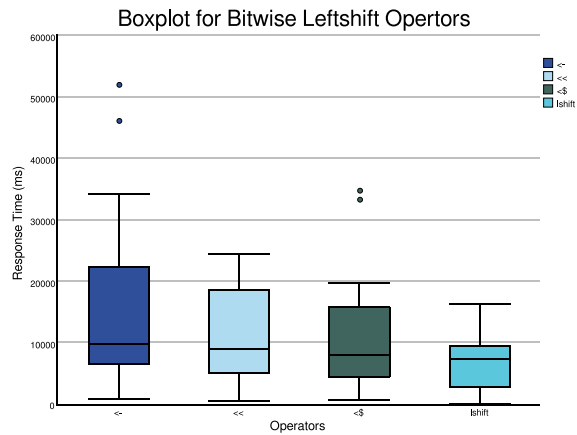}
    \caption{Boxplot Bitwise Leftshift}
    \label{fig:lshiftboxplot}
\end{figure}
\begin{figure}
    \centering
    \includesvg[width=8cm]{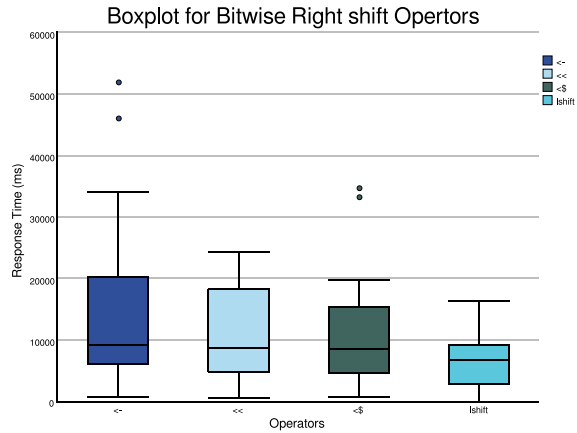}
    \caption{Boxplot Bitwise Rightshift}
    \label{fig:rshiftboxplot}
\end{figure}

\subsection{Analytic Strategy}
The analytic strategy for this study consists of several key steps. First, descriptive analysis will be conducted to extract the mean, standard deviation, and variance of the response time for each operator. Next, a one-way ANOVA test will be employed to analyze the statistical differences among groups of operators, including AND, OR, XOR, NOT, Left Shift, and Right Shift. Subsequently, Tukey's Honestly Significant Difference (HSD) multiple comparison test will be utilized to identify any significant differences between each pair of operators. Finally, multiple linear regression analysis will be performed to examine the relationship between operators and their effect on predicting the dependent variable, which is response time.

\section{Results}
\subsection{Participant Flow}
The study progressed through three phases of development. In the initial phase, data were collected from only one participant, who was the author. In the second phase, to address the limitation of a small sample size, eight participants were included using convenience sampling. These participants were selected from a computer science background. In the third phase, an additional 16 participants were recruited, bringing the total number of participants to 24, excluding the author's data to mitigate bias.

\subsection{Recruitment}
Recruitment for this experiment started after the experiment was designed on February 1, 2024, and ended on May 1, 2024.

\subsection{Statistics and Data Analysis}
Table 3 presents a summary of one-way ANOVA (Analysis of Variance) results for the different operators. The operators considered are AND, OR, XOR, NOT, Left Shift, and Right Shift. In the table, Sum of Squares is denoted as SS, the test statistic as F, degrees of freedom as df, probability as p, and effect size as $\eta^2$.

\begin{table}[htbp]
\centering
\caption{Summary of One-Way ANOVA Results for Different Operators}
\label{tab:anova_results}
\begin{tabular}{|c|c|c|c|c|c|c|c|c|}
\hline
Operator & Sample No. & \multicolumn{5}{c|}{Source: Operators} & \multicolumn{2}{c|}{Source: Residual} \\ \hline
         &            & SS & F & df & p & $\eta^2$ & SS & df \\ \hline
AND      & 5          & 4.75E+08        & 2.105          & 4                   & 0.085       & 0.068                 & 6.49E+09        & 115                 \\ \hline
OR       & 3          & 8.05E+08        & 3.71           & 2                   & 0.029       & 0.097                 & 7.48E+09        & 69                  \\ \hline
XOR      & 4          & 8.78E+08        & 1.113          & 3                   & 0.348       & 0.035                 & 2.42E+10        & 92                  \\ \hline
NOT      & 4          & 6.29E+08        & 4.998          & 3                   & 0.003       & 0.14                  & 3.86E+09        & 92                  \\ \hline
Left Shift   & 4          & 7.94E+08        & 3.086          & 3                   & 0.031       & 0.091                 & 7.90E+09        & 92                  \\ \hline
Right Shift   & 4          & 1.95E+08        & 0.906          & 3                   & 0.441       & 0.029                 & 6.59E+09        & 92                  \\ \hline
\end{tabular}

\end{table}

At the 0.05 significance level, the AND operator does not exhibit statistically significant differences between its means (p = 0.085). The OR operator, however, shows statistically significant differences between its means (p = 0.029), accounting for 9.7\% of the total variance. The differences for the XOR operator are not significant at this level (p = 0.348), explaining only 3.5\% of the variance. The NOT operator demonstrates highly significant differences between its means (p = 0.003), with the largest effect size of 14\% among all operators. The Left Shift operator also has statistically significant differences (p = 0.031), contributing 9.1\% to the total variance. Notably, the Right Shift operator does not have significant differences between its means at the 0.05 level (p = 0.441) and contributes the least to the variance at 2.9\%. In summary, based on the 0.05 significance level and the corresponding p-values, the NOT (p = 0.003) and Left Shift (p = 0.031) operators stand out with substantial and significant differences, while the XOR (p = 0.348) and Right Shift (p = 0.441) operators exhibit insignificant effects on the dependent variable.

\begin{table}[htbp]
  \centering
  \caption{Descriptive Statistics for Bitwise AND Operator Response Time}
    \begin{tabular}{|l|l|l|l|l|}
    \hline
    Operators & N     & Mean      & Std. Deviation & Variance\\ \hline
    $\&$  & 24    & 6816.04   & 4206.568     & 17695215.955 \\ \hline
    $*$ & 24    & 11796.67  & 10602.445    &  112411830.058\\ \hline
    and & 24    & 10904.46  & 7269.235     &  52841780.781\\ \hline
    $\&\&$ & 24    & 11493.17  & 7889.771    &   62248483.797\\ \hline
    $\centerdot$ & 24    & 8117.08   & 6086.726   &   37048233.297 \\ \hline
    \end{tabular}%
  \label{tab:descriptive-stats}%
\end{table}%

As shown in Table 4, the operator '$\&$' exhibits the lowest mean response time of 6816.04 milliseconds, indicating it is generally solved faster than the other operators. The '$*$' and '$\&\&$' operators have similar and significantly higher mean response times of around 11500 milliseconds. Notably, the '$*$' operator has the highest variance (112,411,830.058) and standard deviation (10,602.445), suggesting its response time is more spread out or inconsistent compared to the others.

\begin{table}[htbp]
  \centering
  \caption{Descriptive Statistics for Bitwise OR Operator Response Time}
    \begin{tabular}{|l|l|l|l|l|}
    \hline
    Operators & N     & Mean      & Std. Deviation & Variance \\ \hline
    $\vert$ & 24    & 17182.67  & 15371.210      & 236274102.493 \\ \hline
    or & 24    & 9208.54   & 5762.838       & 33210299.737  \\ \hline
    $\Vert$ & 24    & 11581.54  & 7473.862       & 55858620.433  \\ \hline
    \end{tabular}%
  \label{tab:descriptive-stats-or}%
\end{table}%

Table 5 presents descriptive statistics for the response time of the bitwise OR operators. The operator '$or$' exhibits the lowest mean response time of 9,208.54 milliseconds, indicating it is generally solved faster than the other operators. In contrasting, the '$|$' operator has the highest mean response time of 17,182.67 milliseconds. The '$|$' operator also has the highest variance (236,274,102.493) and standard deviation (15,371.210), suggesting its response time is more spread out or inconsistent compared to the others.

\begin{table}[htbp]
  \centering
  \caption{Descriptive Statistics for Bitwise XOR Operator Response Time}
    \begin{tabular}{|l|l|l|l|l|}
    \hline
    Operators & N     & Mean      & Std. Deviation & Variance \\ \hline
    xor & 24    & 20043.71  & 22159.569      & 491046510.476 \\ \hline
    $\wedge$ & 24    & 15493.58  & 11766.944      & 138460972.949 \\ \hline
    ExclusiveOR & 24    & 16413.25  & 19115.936      & 365419021.326 \\ \hline
    $\wedge\wedge$ & 24    & 11541.12  & 7578.047       & 57426795.158  \\ \hline
    \end{tabular}%
  \label{tab:descriptive-stats-xor}%
\end{table}%

Table 6, it presents descriptive statistics for the response time of the bitwise XOR operators. The operator ‘$\wedge\wedge$’ exhibits the lowest mean response time of 11,541.12 milliseconds, indicating it is generally solved faster than the other operators. In contrast, the '$xor$' operator has the highest mean response time of 20,043.71 milliseconds. The '$xor$' operator also has the highest variance (491,046,510.476) and standard deviation (22,159.569), suggesting its response time is more spread out or inconsistent compared to the others.

\begin{table}[htbp]
  \centering
  \caption{Descriptive Statistics for Bitwise NOT Operator Response Time}
    \begin{tabular}{|l|l|l|l|l|}
    \hline
    Operators & N     & Mean      & Std. Deviation & Variance \\ \hline
    $-$ & 24    & 12141.21  & 9230.495       & 85202032.520 \\ \hline
    $\sim$ & 24    & 6934.83   & 5389.832       & 29050292.058 \\ \hline
    not & 24    & 5822.33   & 4036.011       & 16289383.275 \\ \hline
    $!$ & 24    & 6149.92   & 6112.543       & 37363176.514 \\ \hline
    \end{tabular}%
  \label{tab:descriptive-stats-not}%
\end{table}%

Table 7 presents descriptive statistics for the response time of the bitwise NOT operators. The operator '$not$' exhibits the lowest mean response time of 5,822.33 milliseconds, indicating it is generally solved faster than the other operators. In contrast, the '$-$' operator has the highest mean response time of 12,141.21 milliseconds. The '$-$' operator also has the highest variance (85,202,032.520) and standard deviation (9,230.495), suggesting its response time is more spread out or inconsistent compared to the others.

\begin{table}[htbp]
  \centering
  \caption{Descriptive Statistics for Bitwise LeftShift Operator Response Time}
    \begin{tabular}{|l|l|l|l|l|}
    \hline
    Operators & N        & Mean     & Std. Deviation & Variance \\ \hline
    $<-$ & 24   & 15022.38 & 13474.085      & 181550978.071 \\ \hline
    $<<$ & 24   & 11038.13 & 7580.043       & 57457045.245 \\ \hline
    $<$ & 24   & 11183.25 & 9188.301       & 84424871.152 \\ \hline
    lshift & 24   & 6889.87  & 4453.314       & 19832008.810 \\ \hline
    \end{tabular}%
  \label{tab:descriptive-stats-lshift}%
\end{table}%

Table 8 presents descriptive statistics for the response time of the bitwise Left Shift operators. The operator '$lshift$' exhibits the lowest mean response time of 6,889.87 milliseconds, indicating it is generally solved faster than the other operators. In contrast, the '$<-$' operator has the highest mean response time of 15,022.38  milliseconds. The '$<-$' operator also has the highest variance (181,550,978.071) and standard deviation (13,474.085), suggesting its response time is more spread out or inconsistent compared to the others.

\begin{table}[htbp]
  \centering
  \caption{Descriptive Statistics for Bitwise RightShift Operator Response Time}
    \begin{tabular}{|l|l|l|l|l|}
    \hline
    Operators & N        & Mean     & Std. Deviation & Variance \\ \hline
    rshift & 24   & 11156.67 & 9790.281      & 95849599.797 \\ \hline
    $>>$ & 24   & 8511.79  & 7620.991      & 58079499.824 \\ \hline
    $>\$$ & 24   & 9090.58  & 10203.524     & 104111893.558 \\ \hline
    $->$ & 24   & 7207.04  & 5336.012      & 28473026.216 \\ \hline
    \end{tabular}%
  \label{tab:descriptive-stats-rshift}%
\end{table}%

Table 9 presents descriptive statistics for the response time of the bitwise Right Shift operators. The operator '$->$' exhibits the lowest mean response time of 7,207.04 milliseconds, indicating it is generally solved faster than the other operators. In contrast, the '$rshift$' operator has the highest mean response time of 11,156.67  milliseconds. The '$<-$' operator also has the highest variance (104,111,893.558) and standard deviation (10,203.524), suggesting its response time is more spread out or inconsistent compared to the others.

The table presents the results of Tukey's Honestly Significant Difference (HSD) multiple comparison test for different operators, focusing on the comparison of response times. The results indicate whether the differences in response times between operators are statistically significant. For instance, with the OR operator, the comparison between '$\vert$' and 'or' yields a t-value of 2.65 with a p-value of .027, indicating a statistically significant difference. Similarly, for the NOT operator, there is a statistically significant difference between '$-$' and '$\sim$', '$-$' and '$not$', and '$-$' and '$!$', with probabilities of 0.033, 0.006, and 0.01, respectively. Additionally, the comparison between '$<-$' and '$lshift$' yields a p-value of 0.016, indicating a statistically significant difference. All other comparisons indicate no significant difference.

\begin{figure}
    \centering
    \includesvg[width=15cm]{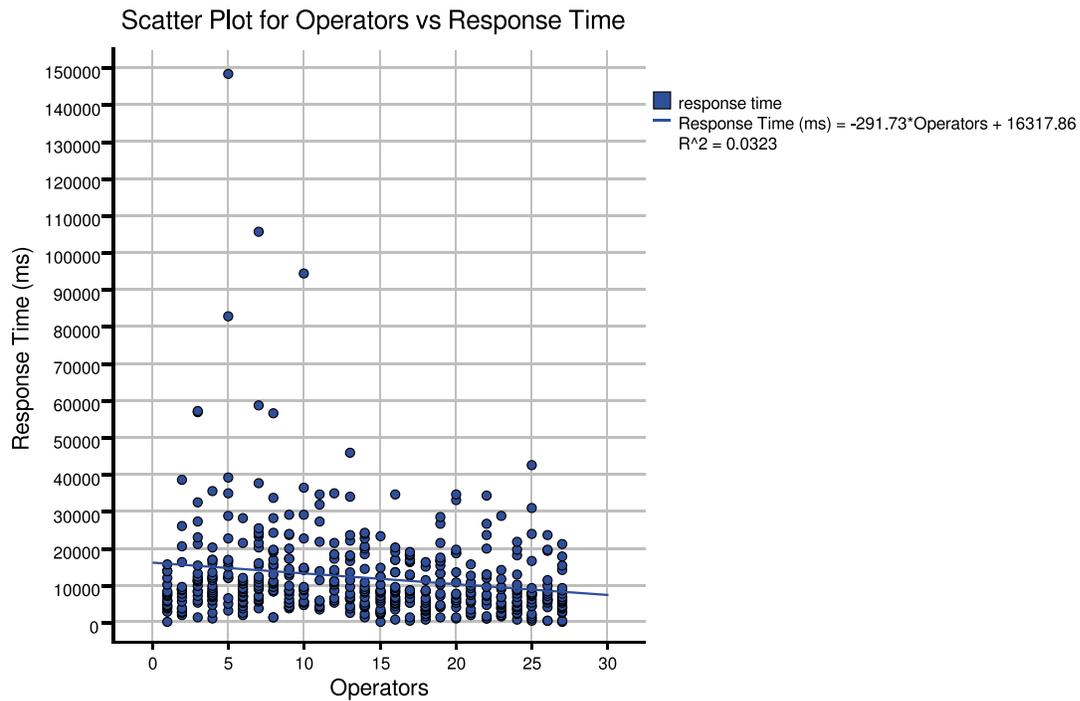}
    \caption{Scatter Plot Operators vs Response Time}
    \Description{This scatter plot shows the relationship between different bitwise operators and the measured response times for the experiment participants. Each operator is plotted along the x-axis with corresponding average response times on the y-axis. The plot demonstrates how the choice of operator affects task completion time.}
    \label{fig:scatterplot}
\end{figure}

As shown in the Figure 9, a multiple linear regression was calculated. Operators appeared to significantly predict response time $R^2 = .032, F(1, 494) = 16.5, p < .001$.

\begin{table}[htbp]
\centering
\caption{Tukey's HSD Multiple Comparison Test Results for Different Operators}
\label{tab:tukey_results}
% \begin{adjustbox}{max width=\textwidth}
\begin{tabular}{|c|c|c|c|c|}
\hline
Operator    & Comparison of Response Time & $t(df)$    & $p$     & Result           \\ 
\hline
AND         & $\&$ vs. $*$     & -2.3(115)  & .154    & Not significant  \\
AND         & $\&$ vs. $and$     & -1.89(115) & .331    & Not significant  \\
AND         & $\&$ vs. $\&\&$     & -2.16(115) & .204    & Not significant  \\
AND         & $\&$ vs. $\cdot$     & -0.6(115)  & .975    & Not significant  \\
AND         & $*$ vs. $and$     & 0.41(115)  & .994    & Not significant  \\
AND         & $*$ vs. $\&\&$     & 0.14(115)  & 1.0     & Not significant  \\
AND         & $*$ vs. $\cdot$     & 1.7(115)   & .44     & Not significant  \\
AND         & $and$ vs. $\&\&$     & -0.27(115) & .999    & Not significant  \\
AND         & $and$ vs. $\cdot$     & 1.29(115)  & .701    & Not significant  \\
AND         & $\&\&$ vs. $\cdot$     & 1.56(115)  & .528    & Not significant  \\
\hline
OR          & $\vert$ vs. $or$       & 2.65(69)   & .027    & Significant      \\
OR          & $\vert$ vs. $\Vert$       & 1.86(69)   & .157    & Not significant  \\
OR          & $or$ vs. $\Vert$       & -0.79(69)  & .711    & Not significant  \\
\hline
XOR         & $xor$ vs. $\wedge$     & 0.97(92)   & .766    & Not significant  \\
XOR         & $xor$ vs. $ExclusiveOR$     & 0.78(92)   & .865    & Not significant  \\
XOR         & $xor$ vs. $\wedge\wedge$     & 1.82(92)   & .273    & Not significant  \\
XOR         & $\wedge$ vs. $ExclusiveOR$     & -0.2(92)   & .997    & Not significant  \\
XOR         & $\wedge$ vs. $\wedge\wedge$     & 0.84(92)   & .833    & Not significant  \\
XOR         & $ExclusiveOR$ vs. $\wedge\wedge$     & 1.04(92)   & .726    & Not significant  \\
\hline
NOT         & $-$ vs. $\sim$     & 2.78(92)   & .033    & Significant      \\
NOT         & $-$ vs. $not$     & 3.38(92)   & .006    & Significant      \\
NOT         & $-$ vs. $!$     & 3.2(92)    & .01     & Significant      \\
NOT         & $\sim$ vs. $not$     & 0.59(92)   & .933    & Not significant  \\
NOT         & $\sim$ vs. $!$     & 0.42(92)   & .975    & Not significant  \\
NOT         & $not$ vs. $!$     & -0.18(92)  & .998    & Not significant  \\
\hline
Left Shift  & $<-$ vs. $<<$ & 1.49(92)  & .448    & Not significant  \\
Left Shift  & $<-$ vs. $<$ & 1.44(92)  & .481    & Not significant  \\
Left Shift  & $<-$ vs. $lshift$ & 3.04(92)  & .016    & Significant      \\
Left Shift  & $<<$ vs. $<$ & -0.05(92) & 1.0     & Not significant  \\
Left Shift  & $<<$ vs. $lshift$ & 1.55(92)  & .412    & Not significant  \\
Left Shift  & $<$ vs. $lshift$ & 1.61(92)  & .381    & Not significant  \\
\hline
Right Shift & $rshift$ vs. $>>$ & 1.08(92)  & .701    & Not significant  \\
Right Shift & $rshift$ vs. $>\$$ & 0.85(92)  & .833    & Not significant  \\
Right Shift & $rshift$ vs. $->$ & 1.62(92)  & .375    & Not significant  \\
Right Shift & $>>$ vs. $>\$$ & -0.24(92) & .995    & Not significant  \\
Right Shift & $>>$ vs. $->$ & 0.53(92)  & .951    & Not significant  \\
Right Shift & $>\$$ vs. $->$ & 0.77(92)  & .867    & Not significant  \\ \hline
\end{tabular}
% \end{adjustbox}
\end{table}

\section{Discussion}

\subsection{Support of Original Hypothesis}
The analysis using multiple linear regression revealed that operators significantly predict response time, with a p-value of less than 0.001 and an R-squared value of 0.032. Therefore, the results suggest that null hypothesis 1 is rejected, although the dataset is small. To present this finding accurately, further investigation is necessary with a larger dataset and more detailed error analysis to draw broader conclusions about this association within bitwise operations.

Regarding Hypothesis H2, the results of the One-Way ANOVA test and Tukey's HSD Multiple Comparison Test indicate that most operators do not show a significant difference in task completion time when compared to each other. A few operators, namely or, not, and left shift, appear to be significant. However, overall, the results support Hypothesis 2, suggesting that the complexity of bitwise operators does not result in longer task completion times.

\subsection{Similarity of Results}
While there are no directly comparable studies available for reference, it is important to contextualize our findings within the broader landscape of research on bitwise operators and programming language development. Our study contributes to this field by examining the intuitiveness of bitwise operators through task completion time and error rates, shedding light on their practical implications for developers.

\subsection{Interpretation}
This empirical study investigated the intuitiveness of bitwise operators through task completion time and error rates. The results highlight that operators can be one of the factors predicting response time. Based on multiple linear regression analysis, we observe a small, negative relation indicating that participants took longer to give a correct answer compared to giving an incorrect one.

The results also indicate that most operators in the group of AND, OR, NOT, XOR, Left Shift, and Right Shift are not statistically significant when compared to each other, as shown by Tukey's HSD Multiple Comparison test, with most comparisons showing no significant differences except for a few operators such as or, not, and Left Shift.

\subsection{Generalizability}
The conclusions of this study are primarily applicable to the community of programming language developers. We consider the factors of operator accuracy and response time to present the intuitiveness of bitwise operators. However, there are multiple factors that may not have been considered, resulting in incomplete reliability of the study. Additionally, given the small number of participants, the reliability of the findings is limited and there are concerns about sampling validity. The results may not be representative of the broader population of programmers or individuals with varying levels of experience. However, expanding this study could provide insights into objectively deriving the intuitiveness of operators.

\subsection{Implications}
Future research should aim to replicate these findings with larger and more diverse samples to enhance the generalizability and robustness of the results. Studies involving participants with varying levels of programming experience would provide a more comprehensive understanding of the impact of bitwise operators on performance metrics.

\bibliographystyle{ACM-Reference-Format}
\bibliography{references}

\end{document}